# Evolution and Development of Brain Networks: From *Caenorhabditis elegans* to *Homo sapiens*

Marcus Kaiser[1, 2, 3*] & Sreedevi Varier[1]

[1] School of Computing Science, Newcastle University, UK
[2] Institute of Neuroscience, Newcastle University, UK
[3] Department of Brain and Cognitive Sciences, Seoul National University, South Korea

**Abstract**
Neural networks show a progressive increase in complexity during the time course of evolution. From diffuse nerve nets in *Cnidaria* to modular, hierarchical systems in macaque and humans, there is a gradual shift from simple processes involving a limited amount of tasks and modalities to complex functional and behavioral processing integrating different kinds of information from highly specialized tissue. However, studies in a range of species suggest that fundamental similarities, in spatial and topological features as well as in developmental mechanisms for network formation, are retained across evolution. 'Small-world' topology and highly connected regions (hubs) are prevalent across the evolutionary scale, ensuring efficient processing and resilience to internal (e.g. lesions) and external (e.g. environment) changes. Furthermore, in most species, even the establishment of hubs, long-range connections linking distant components, and a modular organization, relies on similar mechanisms. In conclusion, evolutionary divergence leads to greater complexity while following essential developmental constraints.

---

[*] Corresponding author: Dr Marcus Kaiser, m.kaiser@ncl.ac.uk

The brain is an extremely sophisticated neural network (Bullmore and Bassett, 2011; Kaiser, 2011; Sporns et al., 2004). The increasing complexity of brain networks coincides with the evolutionary specialization in life forms. Coelenterates such as *Cnidaria* are the first to exhibit neural networks and show a diffuse two-dimensional nerve net, often referred to as a regular or lattice network (Figure 1A). Such lattice networks, with well-connected neighbours and no long distance connections, are a fundamental unit of neural systems, existing even in complex systems like the retina and in the layered architecture of cortical and sub-cortical structures. Sensory organs and motor units require functional specialization and this begins with aggregation of neurons spatially into ganglia or topologically into modules (Figure 1B), as in the roundworm *Caenorhabditis elegans* (Achacoso and Yamamoto, 1992; White et al., 1986). Spatial and topological modules do not necessarily overlap, however both tend to be well connected internally, with fewer connections to the rest of the network. Further up the evolutionary scale, we see greater complexity as in the visual processing system of the rhesus monkey (macaque). Here the visual module consists of two network components: the dorsal pathway for processing object movement and the ventral pathway for processing objects features such as colour and form (Young, 1992). These networks where smaller sub-modules are nested within modules (Figure 1C) are one type of hierarchical network (Kaiser et al., 2010).

As brain size increases, local connections alone, such as for a lattice network become insufficient for integrating information. Brain networks therefore show a small-world organization which not only includes a high degree of connectivity between neighbours but also long-range connections that act as 'short-cuts' linking distant parts of the network. Small-world features are observed in species ranging from *C. elegans* (Watts and Strogatz, 1998) to cat (Scannell et al., 1995), macaque (Hilgetag and Kaiser, 2004), and human (Hagmann et al., 2008), despite different levels of brain size and organization.

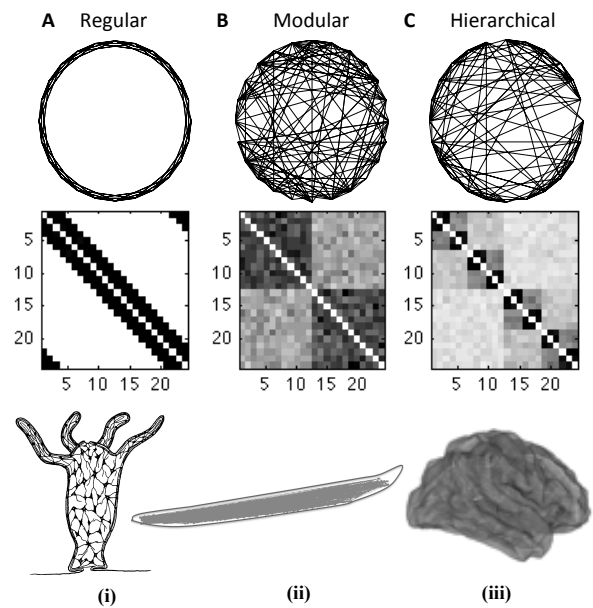

*Figure 1.* Examples for different types of neural networks. **Top row:** (A) Regular or lattice network. (B) Modular network with two modules. (C) Hierarchical network with two modules consisting of two sub-modules each. Each network contains 24 nodes and 72 bi-directional connections (top: circular arrangement placing nodes with similar neighbours close to each other, thus visualizing modules if present in the network; **middle row:** average connection frequency for 100 networks of respective type with colour range from black for edges that occur all the time to white for edges that never occur): **bottom row:** species possessing the afore detailed network architecture (Images are not to scale). (i) Polyp stage of Hydra of the phylum Cnidaria (Image adapted from Ivy Livingstone's drawing in Biodidac) showing a nerve net (ii) Nematode C. elegans showing a modular network (Note that drawing does not take into account the fasciculation of axon tracts) and (iii) Global human neural network traced by Diffusion Tensor Imaging.

Long-distance connections that form short-cuts in a network are expensive in terms of establishment (e.g. myelination and axon guidance) and signal transmission. Development attempts to balance for cost and efficiency, hence, although neural systems in *C. elegans* and macaque tend to reduce the amount of long-distance connectivity (Cherniak, 1994; Chklovskii et al., 2002), studies indicate that re-arranging node positions could reduce wiring length by 50% and 30%, respectively (Kaiser and Hilgetag, 2006). However, a recent study in *C. elegans* indicates that up to 70% of long-distance connections could be formed early during development when *C. elegans* has only a fifth of its adult length (Varier and Kaiser, 2011). As a result, this reduces the need for guidance cues for covering long distances. The significance of long-distance connections is in reducing the average number of intermediate steps within pathways leading to faster information processing, higher reliability, and facilitated synchronization (Kaiser and Hilgetag, 2006). Indeed, a reduced amount of long-distance connectivity was found for disorders that often lead to cognitive deficits as in schizophrenia, epilepsy, and Alzheimer's disease.

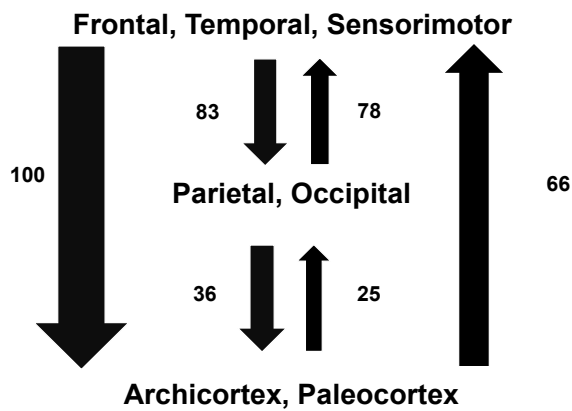

*Figure 2.* Projection patterns, based on tract-tracing studies in the macaque, between different classes of brain regions. Numbers denote the number of projections for each given direction (also indicated by the width of the arrows). The number of projections between frontal, temporal and sensorimotor cortex on the one hand and parietal and occipital cortex on the other hand is almost balanced. However, both classes of cortical regions have more projections to the ontogenetically and phylogenetically older regions of archicortex and paleocortex than projections originating from these earlier maturing regions.

As brain networks evolved to become more complex, there was the inherent need to endow them with greater resilience in the face of injury. Regular networks seen in simpler life forms have a higher degree of redundancy and can therefore cope even with targeted removal of nodes. However redundancy in complex networks is expensive and we see emergence of another feature - hubs - nodes that have significantly more connections than others. Hubs are central in integrating and distributing information and serve to integrate multi-sensory information in brain networks (Sporns et al., 2007; van den Heuvel and Sporns, 2011; Zamora-Lopez et al., 2010). We also see that in newer species, evolutionarily older brain regions tend to contain more highly-connected nodes (Figure 2). Sub-cortical regions such as hippocampus and amygdala are the most highly-connected nodes of the macaque (Kaiser et al., 2007) and occipital and parietal regions show more connections in human cortical networks (Hagmann et al., 2008). While targeted removal of hubs can severely affect network integrity, random removal of nodes will, on average, pick nodes with few connections potentially leading to a smaller deficit after removal (Kaiser et al., 2007). In addition, hubs tend to be connected to each other (van den Heuvel and Sporns, 2011; Zamora-Lopez et al., 2010) leading to an increased resilience towards lesions. Interestingly, in mammals, major hubs are in the centre of the brain, forming early during development.

In conclusion, the network architecture becomes more complex both during development and evolution going from a diffuse lattice organization to hierarchical modular networks. Over time, parts of the network specialize leading to network modules and later to multiple hierarchical levels. There is however an associated cost, namely, the protracted period of brain development and functional maturation needed to achieve the specialization. While behavioural traits like maternal nurture, provide a buffering mechanism, there remains a wider window of vulnerability, when injury can be harder to recover from (Varier et al., 2011). In conclusion, evolutionary divergence leads to greater complexity while following essential developmental constraints, like those influencing hub formation, long-distance connections and modular organization.

**Acknowledgements**
M.K. was supported by WCU program through the National Research Foundation of Korea funded by the Ministry of Education, Science and Technology (R32-10142), the CARMEN e-Science project (www.carmen.org.uk) funded by EPSRC (EP/E002331/1), and (EP/G03950X/1). S.V. was funded through EPSRC (EP/G03950X/1).